\def\ltsim{\mathrel{<\kern-1.0em\lower0.9ex\hbox{$\sim$}}}
\def\gtsim{\mathrel{>\kern-1.0em\lower0.9ex\hbox{$\sim$}}}
\def\app{$\approx$}
\def\tende{\rightarrow}
\documentclass{aa}
\begin{document}
\title{ Multi--frequency Analysis of the two CSS quasars \\ 3C 43 \& 3C 298}
\author{{C. Fanti} \inst{1,2} \and {R. Fanti} \inst{1,2} \and {D. Dallacasa}
\inst{1,3} \and A. McDonald, \inst{6} \and {R.T. Schilizzi} \inst{4,5}
\and {R.E. Spencer} \inst{6}}
\offprints{D. Dallacasa}
 \mail{ddallaca@ira.cnr.it}

\institute{{Istituto di Radioastronomia del CNR, via Gobetti 101, I--40129
Bologna, Italy} 
\and {Physics Dept., University of Bologna, via Irnerio 46, I--40126
Bologna} \and {Astronomy Dept., University of Bologna, via Ranzani 1, I--40127
Bologna, Italy } 
\and {Joint Institute for VLBI in Europe, Postbus 2, 7990 AA, Dwingeloo, The
Netherlands} \and {Leiden Observatory, Postbus 9513, Leiden, 2300RA, The Netherlands}
\and {University of Manchester, Jodrell Bank Observatory, UK}}
\date{Received {26 June 2002}  / Accepted 24 September 2002}
%
\markboth{3C 43 \& 3C 298}{Dallacasa ed al.}
\titlerunning{3C 43 \& 3C 298}
\authorrunning{C. Fanti et al.} 
\abstract{We present and discuss observations made with MERLIN and VLBI at
1.7 and 5 GHz of the two CSS quasars 3C~43 and 3C~298. They show quite
different morphologies, the former being a very distorted triple radio
source, the latter a small FRII type object. Relativistic effects and  structural
distortions are discussed.  Source ages  are
evaluated to be of the order of $\approx 10^5$ years, therefore 3C~43 and
3C~298 can be considered  fairly ``young'' radio sources.
Some inference is also derived on the properties of the medium  
surrounding the radio emitting regions in these sub--galactic objects,  whose density could be
as low as 10$^{-3}$ cm$^{-3}$. 
}
\maketitle

\section{Introduction}
    Compact Steep--spectrum Sources ({\it CSS}) and GHz Peaked spectrum Sources
    ({\it GPS}) are powerful objects whose projected sizes are
    shown, statistically (Fanti et al. \cite{fan3}), to be {\it physically small}
    ($<15-20~h^{-1}$ kpc)\footnote{$H_-1=100h$ km/sec/Mpc, $q_0 = 0.5$}. 
    Their high--frequency spectrum is steep (and turns
over at low frequencies in {\it GPS}) implying that these objects are {\it
not} core dominated. When observed with the appropriate resolution they
display a large variety of morphologies.
    Their nature has been a matter of debate for many years and several samples 
    of CSSs and GPSs have been studied by numerous authors in an effort to
understand
their properties and their r\^ole in the radio source evolution. General
    discussions have been presented, for
    instance, by Fanti et al. (\cite{fan1}), Saikia (\cite{saik}), Fanti et al.
(\cite{fan3}), Spencer et al. (\cite{spen2}), 
Fanti et al. (\cite{fan5}), Readhead et al. (\cite{read}), O'Dea \& Baum (\cite{ode2}).
    An extensive review has been presented by O'Dea (\cite{ode}).

It is now generally accepted that at least a large fraction of
CSSs/GPSs are  young radio sources in the early
stage of their life ($\ltsim 10^6$ years; see e.g. Fanti \cite{fan6} for a short
review) and statistical studies of CSSs and GPSs are therefore important in
order to refine the 
evolutionary scenario. Extensive studies of individual objects 
are also important, however, in order to understand the underlying physics.
Moreover, due to their small physical sizes, these objects may give us an unique 
opportunity to probe the interstellar medium (ISM) of their host galaxy/quasar and, in the 
smallest of them, even the Narrow Line Region ({\it NLR }) via jet--ambient gas interactions (see for
instance de Vries, \cite{devr}, Conway \& Schilizzi, \cite{con2}, Axon et al. \cite{axon} and 
references therein, Morganti et al. \cite{morg}).

In this paper we present a detailed  study of  the CSS
    quasars 3C~43 and 3C~298 based on different resolution images 
we obtained using MERLIN and VLBI at 1.7 and 5 GHz and complemented by images at other frequencies.
These quasars have similar redshift and radio power but very 
different radio morphologies: 3C~43 shows a very distorted structure; 
3C~298, instead, has an 
almost     linear morphology and appears to be a scaled down version of the large 
size quasars.

In the following sections we summarize the observations and the data reduction 
(Sect.~\ref{observ}),  present the observational  results 
(Sect.~\ref{data_an}) and discuss the properties of both sources 
(Sect.~\ref{disc}). Finally, in Sect.~\ref{summ} we provide a summary 
of the results and some conclusions.

\section{Observations and Data reduction}
\label{observ}
\begin{table*}[htbp]
\begin{center}
\caption{Observational information for 3C~43 and 3C~298 \hfill}
\medskip
\begin{tabular}{rccr|l}
\hline
 source & $\nu$(GHz) & date    & Obs. Mode&\qquad\qquad           Stations               \\
\hline
 3C 43  & 1.7   &1993.60  & MERLIN   & Jodrell(MK2), Darnhall, Defford, Knockin,
Tabley, Wardle, Cambridge,            \\
       &       &1986.75 & MKII      & Onsala, Medicina, Defford, Effelsberg,
Westerbork, Lovell  \\

       & 5.0   &1991.70 & MERLIN   &Jodrell (MK2), Knockin, Darnhall, Defford, Tabley         \\
       &       &        & MKII      & Effelsberg, Westerbork, Jodrell(MK2), Knockin, Cambridge               \\
       &       &        & MIIIB     & Effelsberg, Westerbork, Jodrell(MK2), Onsala    \\
 \hline
 3C 298 & 1.7   & 1982.82 & MERLIN   & Jodrell(MK2), Darnhall, Defford, Knockin, Tabley, Wardle \\
       &       & 1991.88 & MKII      & Effelsberg, Westerbork, Torun, Crimea, 
       Lovell, Medicina, \\
       &       &         &          & Green Bank, Y27, OVRO, VLBA\_(NL, FD,
LA, PT, KP)                    \\
       & 5.0   & 1983.66 & MERLIN   & Jodrell(MK2), Knockin, Darnhall, Defford, Tabley         \\
       &       & 1992.24 & MKII      & Cambridge, Effelsberg, Jodrell(MK2), Knockin, Medicina\\
       &       &         & MKIIIB     & Effelsberg, Jodrell(MK2), Medicina, Noto, Onsala, Westerbork            \\
\hline
  
\end{tabular}
 \label{obse}
\end{center}
\end{table*}

The new observations presented  in this paper were obtained  with VLBI networks 
in different observing modes  (MkII and MkIII) and with MERLIN, in the period
1991--1994 at the frequencies of 1.7 and 5 GHz (Table~\ref{obse}). 

At 5 GHz the observations of 3C~43 were performed simultaneously with MERLIN and EVN
while for 3C~298 we obtained EVN observations only.
The EVN observing time was 12 hours for each source and
included short scans on the calibration sources 0133+476,  0235+164
 and OQ208, about  every four hours.

At 1.7 GHz we performed for 3C 298 a global ($EVN+US+VLBA$) observation.
The observing time was about 12 hours
 on both  the European and the US networks with only four hours in
 common to the two networks due to the low source  declination.
 For 3C 43 the EVN data by Spencer et al.  (\cite{spen2})
were re-analyzed and a new image is presented here.

Given the complex structure of these sources the use of the combined VLBI and 
MERLIN data
was necessary in order to obtain a good sampling  of both  the short and the 
long baselines. As said above, only for 3C~43 at 5~GHz we did manage to have simultaneous 
observations on the two arrays. The other MERLIN data were: for 3C~43 at 1.7 GHz 
from observations performed in 1993 with the extended array (i.e.
including the 32-m telescope at Cambridge); for 
3C~298 pre--existing data by Spencer at al.  (\cite{spen}) at 1.7 GHz
and by Akujor et al. \cite{aku2} at 5~GHz.
 
In order to combine the VLBI and the MERLIN data sets properly, at least 
one common baseline is highly desirable to link the phases and 
the flux  density  scales. This was not possible for 3C~298 at 1.7 GHz.
In this case the MERLIN and VLBI data were combined after carefully
 checking that the ``a--priori"  flux  density scales of the two arrays
 were in agreement.
 
The whole data reduction was made in AIPS. The final combined images were
 obtained by initially mapping and phase self-calibrating the short baseline
 data and then by slowly adding increasingly longer baselines. 
For both sources we analyzed images made at different
resolutions; they are 
referred to in the text as {\it low, intermediate} and {\it  high} (see 
Tables~\ref{char_43} and \ref{char_298}).

\section{Results}
\label{data_an} 

Some basic parameters for the two quasars 3C 43 and 3C~298 are given in 
  Table~\ref{info}.

\tabcolsep 0.08cm
\begin{table}[htbp]
\begin{center}
\caption{Basic information for 3C 43 and 3C 298 \hfill}
\begin{tabular}{lrrrrrrrr}
\hline
\hline
\ \ source & $z~~$ &  m$_{\rm v}$ & S$_{\rm 1.7}$&S$_{\rm 5.0}$&\ logP$_{1.7}$ &  LS\ \ \ \ \ \   &$\nu_{\rm max}$&$\alpha$ \\
      &     &         &  (Jy)&(Jy)&     & (kpc $~h^{-1}$)& MHz       & 
     \\ 
\hline
\hline
&&&&&&&& \\
3C  43  & 1.46~& 20.0  &2.6~ &1.1$^a$  &  27.83       &$\approx 15\ \ \ \
\ $  &$<$30~~    & 0.71   \\
0127+233&&&&&&&& \\
3C 298  & 1.44~& 16.8 &4.9$^a$& 1.5$^b$  & 28.24        & 6.5\ \ \ \ \ &   80~~  & 1.10  \\ 
1416+067&&&&&&&& \\ \hline\hline
\end{tabular}
    \label{info}
\end{center}
$z$: redshift; m$_{\rm v}$: visual magnitude; ${\rm S}_{5.0}, {\rm S}_{1.7}$:
total flux density at the quoted frequency; ${\rm P}_{1.7}$ in W/Hz $h^{-2}$; LS: overall linear size;
$\nu_{\rm max}$: observed spectral turnover frequency; $\alpha$: overall spectral index 
for $\nu\gtsim$100 MHz, defined as S$(\nu)\propto\nu^{-\alpha}$

$a$ -- Spencer et al. (\cite{spen}); $b$ -- NED
\end{table}

The two sources have been analyzed in similar ways following this scheme:

\begin{figure*}
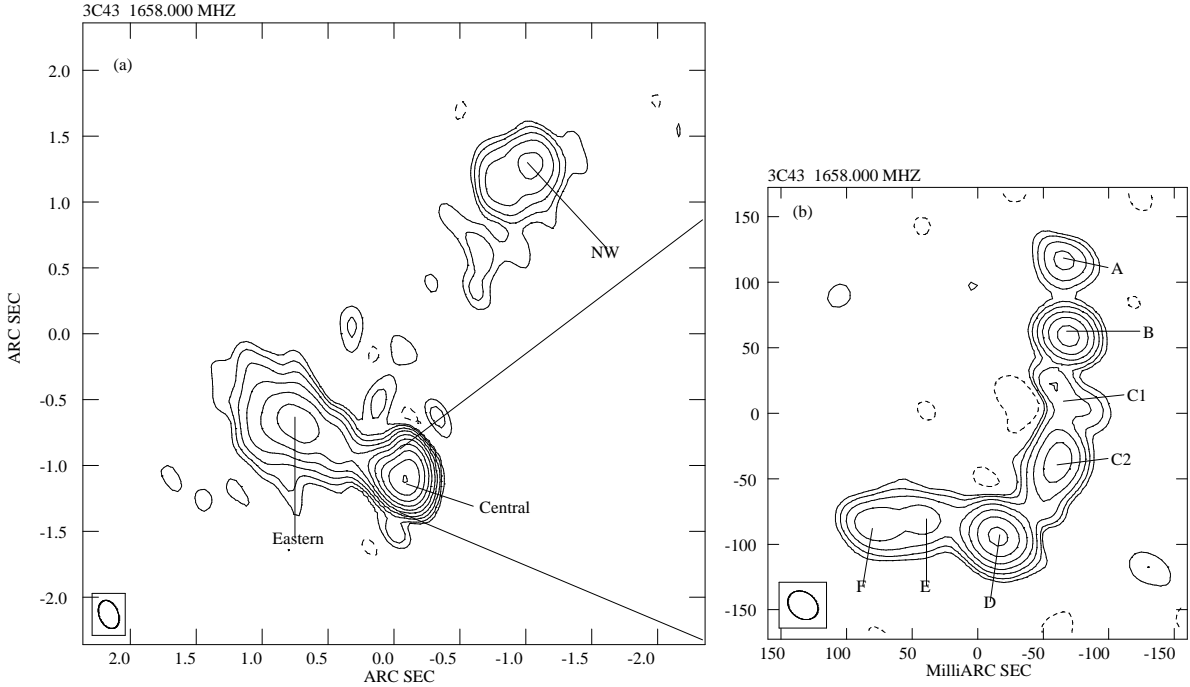

 \vskip 9truecm
\includegraphics{H3874F1A.PS}
\includegraphics{H3874F1B.PS}
 \caption{3C 43 at 18 cm: {\it (a)} MERLIN image ($220 \times 145$mas);
 contours: ($\pm 2\times 2^n, n\geq 0$)  mJy/beam;  $S_{\rm peak}=$1.06 Jy/beam;
{\it (b)} EVN image ($25\times 20$mas) of the
``Central''  component or {\it bright jet}; contours: ($\pm 4.0\times 2^n, 
n\geq 0)$ mJy/beam; $S_{\rm peak}=$0.308 Jy/beam}
\label{new_M}
 \end{figure*}

\begin{itemize}

\item[{\it a)}] The source morphology was studied by
comparing the images of the present paper with others available in the
literature. For the individual components of each source we report in 
Tables~\ref{com43} and \ref{com298} flux density and beam--deconvolved sizes 
(HPW), derived using the AIPS task IMFIT.
When components cannot be reliably approximated by
two--dimensional Gaussians, flux densities have been measured by 
integration over the emission region (AIPS task TVSTAT) and sizes may have
been estimated from the lowest reliable contour in the image.  Note that in this
case sizes are roughly twice the conventional HPW. Parameters estimated in
such a way are preceeded by ``$\sim$''.

\item[{\it b)}] On the assumption of minimum energy conditions,
we computed for each source component the physical parameters,
i.e.: equipartition magnetic field (H$_{\rm eq }$), minimum
energy density (u$_{\rm min}$), minimum total energy (U$_{\rm eq}$), 
turnover frequency ($\nu_{\rm to}$) expected from
synchrotron self--absorption. We used standard formulae (e.g. 
Pacholczyk, \cite{pach}), with filling factor 1, equal energy in electrons
and protons and an ellipsoidal geometry for sub--components.

\item[{\it c)}] Component radio spectra are derived by combining our 
own data with those from images  in the literature at resolutions not too 
different  from ours. 
\end{itemize}

\subsection{3C 43}
    \label{3c43}

\subsubsection{Source Morphology -- Observed and Physical Parameters}
 \label{mor_43}
\begin{figure}[h]
 \vskip 7truecm
\includegraphics{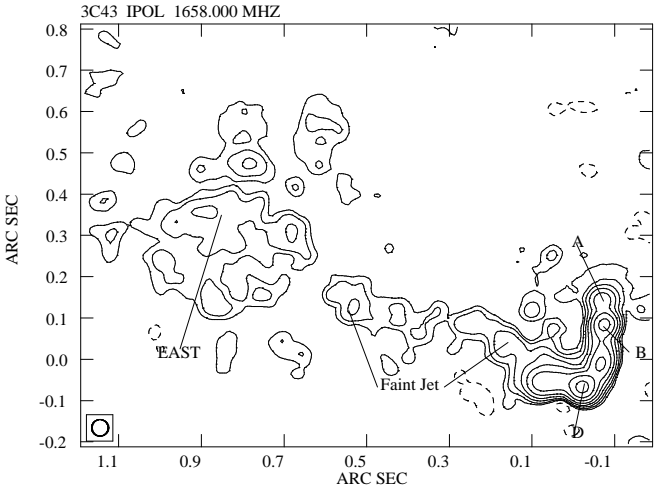}
\caption{3C 43: EVN+MERLIN image at 18 cm of the ``Central" and ``Eastern'' 
[$EAST +$ {\it faint jet}] components  ($40\times 40$mas); contours: 
($\pm 2.5\times 2^n, n\geq 0)$
mJy/beam; $S_{\rm peak}=$0.431 Jy/beam   }
\label{evnM:18:43}
 \end{figure}

 At sub--arcsec resolution (Pearson et al. \cite{pear}; Akujor et al. 
 \cite{aku1}, Spencer et al. \cite{spen}; Akujor \&  Garrington \cite{aku3})
the morphology is that of a very distorted triple  source, 
with a very bright central component and two outer components,
toward East and North--West, whose  arms form an angle of $\approx 95^\circ$.
In the 18 cm image obtained with the extended MERLIN (Fig.~\ref{new_M}a)
the north--western component is clearly double and
the component to East is elongated towards the central one and connected to it.
At higher resolutions  (van Breugel et 
al. \cite{vanb}, L\"udke et al. \cite{lud}) the latter  component appears as 
a faint  jet
which widens at the eastern end in a sort of lobe (see also Fig.~\ref{evnM:18:43}). 
No compact features which could be interpreted as ``hot--spots'' are visible
within the eastern or north--western lobes at either frequency, even
in the {\it high} resolution EVN images.

In the discussion of the source properties we refer to the labelling of Figs.
\ref{new_M} and \ref{evnM:18:43}. In these figures: $NW$, ``Eastern'' and ``Central''
are the north-western, the eastern and the central components seen at
MERLIN resolution (Fig.~\ref{new_M}a); $EAST$ and {\it faint jet} are the
easternmost lobe and the elongated weak structure which connects it to the ``Central''
component (Fig.~\ref{evnM:18:43}); {\it bright jet} is the ``Central''
component as seen at VLBI resolutions (Figs.~\ref{new_M}b and \ref{evn:6:43}).
Here components are labelled from $A$ (North) to $F$ (South--East).

The general parameters  of our own {\it high} and {\it low} resolution VLBI 
images  at 1.7 (Figs.~\ref{new_M}b and \ref{evnM:18:43}) and at 5 GHz 
(Fig.~\ref{evn:6:43}) are given in Table~\ref{char_43}. 

\begin{table}[htbp]
\begin{center}
\caption{Image characteristics for 3C 43 \hfill}
\medskip
\begin{tabular}{l|ccc|ccc}
\hline
&  \multicolumn {3}{c|}{1.7 GHz} & \multicolumn {3}{c}{5 GHz} \\ \hline
resol. &   beam       & Flux$^{\ddag}$ & 
noise$^{\dag}$  &  beam       & Flux$^{\ddag}$ & noise$^{\dag}$  \\
       &    (mas)      &  (Jy)   &(mJy/b)
       & (mas)       &  (Jy)   &(mJy/b)\\
\hline
{\it  high}  &  25$\times$20  & 1.07   & 1.3  &15$\times$10  &  0.49  &  1.0  \\
 {\it low}   & 40$\times$40 &  2.48  &  1.2   &25$\times$17 &   0.88  &  0.9  \\
\hline
\end{tabular}
\label{char_43}
\end{center}

$\dag$  noise computed far from the source

${\ddag}$ flux density in the image; at 1.7 GHz {\it low} resolution
it includes lobe $NW$.
The difference in flux density, at both frequencies, between the {\it high} and
{\it low} resolution images is due to the missing short (MERLIN) baselines in the 
former case, which causes loss of extended low brightness features (see Figs.
\ref{new_M}, \ref{evnM:18:43}, \ref{evn:6:43})

\end{table}

 \begin{figure}[h]
 \vskip 8 truecm
\includegraphics{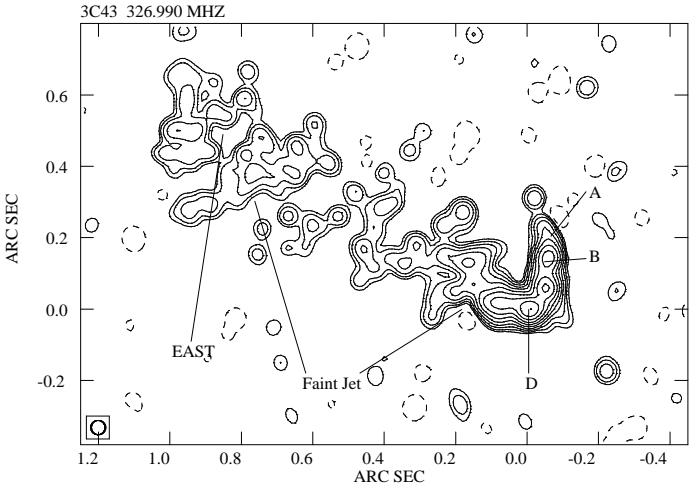}
 \caption{3C 43: EVN image at 92 cm ($40\times 40$mas) of the same region as
in Fig.~\ref{evnM:18:43}; contours: ($\pm 2.0\times 2^n, n\geq 0)$
mJy/beam; $S_{\rm peak}=$1.025 Jy/beam    }
 \label{evn:90:43}
 \end{figure}

The {\it low} resolution image at 1.7 GHz shown in Fig.~\ref{evnM:18:43}
is the only one where  the {\it faint jet} is clearly visible. Component  $NW$ (out of figure) 
is instead just barely detected.
This image accounts for $\approx$95\% of the total MERLIN flux  density.
A global VLBI image at 327 MHz of the ``Central'' and  ``Eastern''  components
(component $NW$ is heavily resolved out), with a resolution of $40\times 40$ 
mas (Dallacasa et al. unpublished)  is shown in Fig.~\ref{evn:90:43} 
for comparison.  Most of
the structure visible in Fig.~\ref{evnM:18:43} can be recognized here.

In the {\it high} resolution images presented in Figs. {\ref{new_M}b 
and~\ref{evn:6:43}a, only the ``Central'' component is visible. It  appears as a 
{\it bright jet} 
running initially  in the N--S direction, with small oscillations and then  
sharply bent towards East (see also the 610 MHz image by Nan et al. 
\cite{nan2}, resolution  $30\times 20$ mas).
Only $\approx 55$\% of the flux density measured at 1.7 GHz with  MERLIN 
and  $\approx 50$\% of that measured at 5 GHz with the VLA by Spencer et al. 
(\cite{spen}) is present in these images. More flux density is detected in the
combined EVN$+$MERLIN image at 5 GHz shown in Fig.~\ref{evn:6:43}b, thanks
to the better coverage of the short baselines. Here $\approx$90\% of the VLA 
flux  density of the  ``Central" component (Spencer et al.  \cite{spen})
has been recovered.

The 5 GHz {\it low} resolution image (Fig.~\ref{evn:6:43}b) compares
reasonably well with the image at 1.7 GHz (Fig.~\ref{new_M}b), with a similar 
 resolution,   although the match is not completely satisfactory.
In particular the extended structure $C1$ appears more
 ``knotty" here and  components $E$ and $F$ are now blended into 
 a  single blob. 
 We note that component $A$ (Fig.~\ref{new_M}b), brighter at high frequencies,
is just barely visible at 327 MHz,  heavily blended with component $B$.  We estimate 
that its flux density is not greater than \app 20 mJy at this frequency. 
This provides evidence that it has an 
inverted spectrum (Sect.~\ref{spec_43}) and that it is therefore likely to 
harbor the  source core, as suggested already by Nan et al. (\cite{nan2})

\begin{figure*}
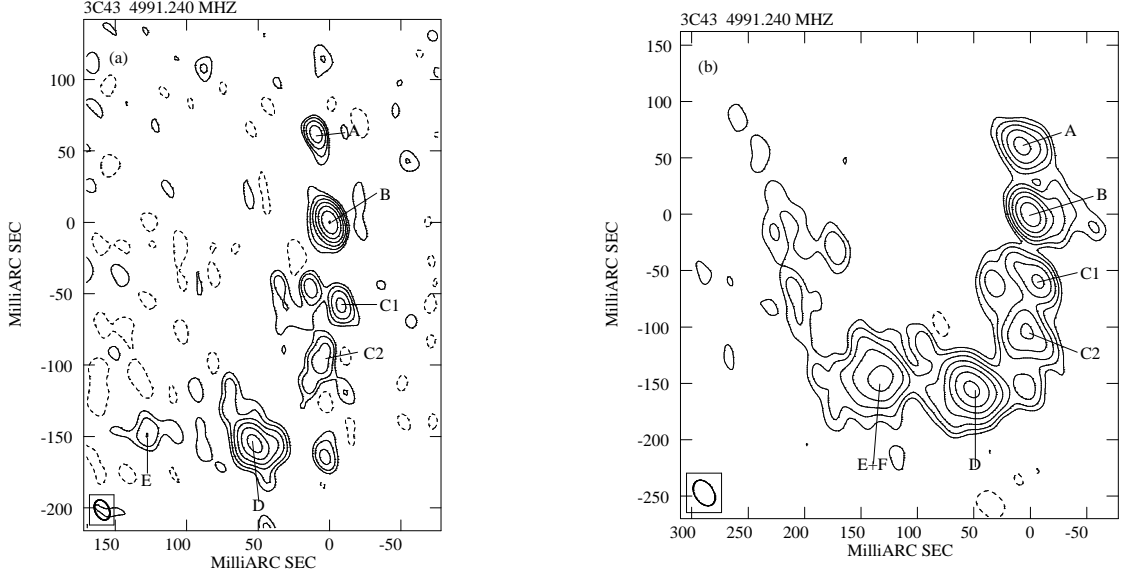

\vskip 9truecm
\includegraphics{H3874F4A.PS}
\includegraphics{H3874F4B.PS}
 \caption{the {\it bright jet} of 3C 43 at 6 cm: {\it (a)} EVN image  
 ($15\times 10$mas); contours:  ($\pm 2.0\times 2^n, n\geq 0)$ mJy/beam; 
 $S_{\rm peak}=$0.131 Jy/beam;  {\it (b)}  EVN+MERLIN image 
 ($25\times 17$mas); contours: ($\pm 2.0\times 2^n, n\geq 0)$
mJy/beam; $S_{\rm peak}=$0.158 Jy/beam     }
 \label{evn:6:43}
 \end{figure*}

\begin{table*}[htbp]
\caption{Observed and Derived  Parameters for 3C 43 \hfill}
\medskip
\begin{tabular}{c|r|rcc|rrrrr}
\hline
& 1.7 GHz & \multicolumn {3}{c|}{5 GHz} & \multicolumn {5}{c}{Physical
Parameters} \\ \hline
comp & S  & S & $\theta_1\times\theta_2$ &  d$_1\times$d$_2$ &
$\alpha_{\rm thin}$ & H$_{\rm eq}\ \ $ & u$_{\rm min}\ \ \ \ $  & U$_{\rm eq}\ \ $ &
$\nu_{\rm to}$\\
     &  mJy &   mJy   &     mas                  & pc $~h^{-1}$            &
                   &   mG         &\  erg/cm$^3~h\ \ $  &~~~ 10$^{54}$erg $h^{-2}$   &
   MHz \\
\hline
  A& 43  &  61 & 17$\times$11 &   73$\times$47  &&&&& \\
       &--  & 26&unres.  &   &   &   &  &  &    \\
  B   &202&121 &  7$\times$6  &  30$\times$26  &0.54
&7.8& 5.7$\times10^{-6}$&10~~~~~& 436\\
    &-- &152 &  8$\times$3  &         &     &   &       &    &    \\
C1     & $\sim$90 & $\sim$65 & 39$\times$21 &  167$\times$90  &0.30
&1.8&3.0$\times10^{-7}$&36~~~~~& 81\\
    --& & $\sim$40& 10$\times$4  &           &     &   &       &    &    \\
C2     & $\sim$180&  $\sim$80 & 30$\times$23 &  129$\times$99  &0.63
&2.4&5.2$\times10^{-7}$&57~~~~~& 142\\
 &-- &  $\sim$30& 22$\times$6  &         &     &   &       &    &    \\
D     &459& 275 & 20$\times$16 &   86$\times$69  &0.48
&4.0&1.5$\times10^{-6}$&54~~~~~& 252\\
  &-- & 175 & 15$\times$10 &          &     &   &       &    &    \\
E+F     &$\sim$200& 108 & 28$\times$26 & 120$\times$111 &0.66 &2.4
&5.1$\times10^{-7}$&68~~~~~& 145\\
 &--  & --       &--   &  --    &     &   &   &  &  \\\hline
  $NW\dag$ &$\sim$185&  50 &400$\times$300  &(1.7$\times1.3)\times 10^3$ &$\sim$ 1.1
&0.4&1.5$\times10^{-8}$& 3600~~~~~&  32\\
  ``Eastern''$\dag$&$\sim$560&  130&600$\times$300\quad &(2.5$\times1.3)\times 10^3$
&1.27&0.6&
  3.5$\times10^{-8}$  &13000~~~~~&48\\ \hline

\end{tabular}
\medskip
\label{com43}

At 5 GHz data on the first line are from the {\it low} resolution image, the 
others from the {\it high} resolution image.

$\dag$ Data at 5 GHz from Spencer et al. (\cite{spen});

$\alpha_{\rm thin}$ spectral index in the frequency range 0.3--5 GHz;
H$_{\rm eq}$ equipartition magnetic field; u$_{\rm min}$ minimum energy
density; U$_{\rm eq}$ minimum energy; $\nu_{\rm to}$ computed self--absorption
turnover frequency (except for $A$)
\end{table*}

The EVN observations at 1.7 GHz were performed in September 1986 and those at
 5 GHz in September 1991: this represents a time lag long enough to 
impose the exercise of  checking if
the source structure has somehow changed. To do this, we have 
convolved the 6 cm {\it low} resolution image to the resolution of the 18 cm
 (25$\times$20 mas; not shown).
We find that all components from $B$ to $D$  appear to have systematically
moved away from $A$ (assumed stationary) along PA$\approx  -140^\circ$,
the positional shifts  being in the range 3.8--12.5 mas.
These displacements are significant and,  taken at face value,
would imply an apparent outward  speed $\beta_{\rm app}  \geq 22 c$.

Such displacements may be explained as a sort of ``reflex motion'' due to
spectral index and resolution effects within component $A$.
In effect $B$ and $D$, the two strongest and most reliable components, have
 not moved appreciably with respect to each other. 
If we then take  as coordinate origin the position of $B$ instead of $A$,
we find that the displacement of $A$ with respect to  $B$  increases
systematically  when we use the
18 cm {\it high} resolution, the 6 cm {\it low} resolution and 
the 6 cm {\it high} resolution images. This systematic
behaviour is consistent with the assumption that $A$ be actually composed
by an inverted spectrum compact ``true core" (dominating at 6 cm {\it high}
resolution)  at the eastern end of a mini--jet with a normal
spectrum, pointing towards $B$ (dominating at 18 cm). 
Due to the different spectral indices, the relative weight
of these two components changes depending on the observing frequency
and resolution, and the position of $A$ shifts with respect to $B$ in a way 
consistent with what observed. 
Such a sub--structure, confirmed by a recent unpublished  VLBA image at 
8.4 GHz by Mantovani (private communication), could help in explaining some
of the points discussed in Sect. \ref{dist3c43}.

Finally, by comparing the VLA polarization information at 5 GHz (Akujor et al.
\cite{aku1}), 8.4 GHz
(Akujor \&  Garrington, \cite{aku3}) and 15 GHz (van Breugel et al.
\cite{vanb})
we find
that both the ``Eastern'' and the ``Central'' components show a fair amount of
depolarization
between 15 and 5 GHz, while the $NW$ component is not depolarized 
between 8.4 and 
5 GHz. Faraday rotation does not seem to be important. The high resolution 
polarization images by L\"udke et al. (\cite{lud}) and by van Breugel et al. 
(\cite{vanb}) show that 
the magnetic field is well aligned parallel to the {\it bright} and the {\it faint
jet}  and follows the bend at $D$.

\smallskip
Observed component parameters are given in Table~\ref{com43}.
 At 1.7 GHz no sizes are given for components $A$ to $F$ since they are not significantly different
from the old ones by Spencer et al. (\cite {spen2}).
At 5 GHz we give flux densities and sizes from 
 both the {\it low} and  the {\it high} resolution images, except for the
 $NW$ and 
``Eastern''  components, whose data  are from Spencer et al. (\cite{spen}).
The derived physical parameters are also reported in Table~\ref{com43}, except for
$A$ which is unresolved with an inverted spectrum (Sect.~\ref{spec_43}).

\subsubsection{Spectral Analysis}
 \label{spec_43}

The overall spectrum of 3C 43 derived from low resolution data (Kuhr et al. \cite{kuh}; Steppe et al. 
 \cite{step}) is straight with $\alpha=$0.71  from $\approx$30 MHz  to 230 GHz.

The addition to the present data of the measurements at 610 MHz (Nan et al. 
\cite{nan2}) and 327 MHz (Dallacasa et al. unpublished) shows that the
spectrum of
the ``Central'' component, or {\it bright jet}, is straight
($\alpha \approx 0.6 \pm 0.05$) at least down to $\approx$ 0.3 GHz. Here some
flattening might be occurring since its flux density is $\approx 14 \%$ lower than that extrapolated from
the higher frequencies.
For the $NW$ and ``Eastern'' ($EAST$ plus {\it faint jet}) components,
data at sub--arcsec
resolution are available at four frequencies (1.7 from Spencer et al.
\cite{spen} and from this paper, 5 and 15 GHz from Spencer et al.  
\cite{spen}, 15 GHz from van
Breugel et al. 1992, and 8.4 GHz from Akujor \&  Garrington, \cite{aku3}).
The spectrum of  the ``Eastern'' component 
is straight and steep ($\alpha \approx 1.3$). The spectrum of component $NW$
is more uncertain, but very likely has  $\alpha \approx 1.1$.
In any case the combined spectrum  of [``Eastern'' $+~ NW$]
has to flatten, at $\leq$ 100 MHz, otherwise the extrapolated 
flux density  would exceed the overall source flux density at low frequencies.

In order to analyze over a broader frequency range 
the spectrum of the extended features (including low surface
brightness ones possibly missed by the present observations)
and search for a frequency break, 
we have subtracted the spectrum of the ``Central'' component from the source 
total spectrum.
The assumption that the spectrum of the ``Central'' component is straight 
and that the flux density missing at 327 MHz is caused by a poor $uv$ coverage
at this frequency, sets a frequency break at $\approx$~300 MHz in the 
{\it subtracted spectrum}.
If, on the contrary, the spectrum of the ``Central'' component 
turns over at $\approx300$ MHz the {\it subtracted spectrum} is well
 fitted by a power law with $\alpha = 1.15$ down to $\approx 100$ MHz.

The knots in the {\it bright jet} compare reasonably well with each other in
the VLBI images at the four available frequencies of 0.3, 0.6,
 1.7 and 5 GHz and it is possible to derive their individual spectral indices
  in this frequency range. They  are given in Table~\ref{com43}
($\alpha_{\rm thin}$).
All components but $A$ are transparent down to 327 MHz and their spectra are
perhaps mildly 
steepening downstream the jet, except at $D$, where the jet sharply bends. 
Their synchrotron self-absorption frequencies, $\nu_{\rm to}$, computed under 
equipartition assumptions (see Table~\ref{com43}), are consistent with 
this finding. 

 The spectrum of $A$ is instead markedly inverted between 0.3 and 5 GHz, with
 $\alpha_{\rm thick}\approx 0.4$; this is a  strong indication of the core location. This
 spectrum must however have a maximum below 15 GHz, since at this frequency
  the datapoints of the total spectrum (Kuhr et al. \cite{kuh}; Steppe et al. 
  \cite{step}) follow a power law with no indication of flattening or bending.
Unpublished VLBA data at 8.4 GHz (Mantovani private communication) indicate
indeed that the spectral peak occurs between 5 and 8.4 GHz.

\subsection{3C 298}
    \label{3c298}

 \subsubsection{Source Morphology -- Observed and Physical Parameters}
 \label{mor_298}

There is a considerable amount of data in the literature on this source at many 
frequencies and resolutions. Images at sub--arcsec resolution, many of which
also have polarization measurements, have been presented
e.g. by Pearson et al. (\cite{pear}); Spencer et al. (\cite{spen}), 
van Breugel et al.
(\cite{vanb}), Akujor \&  Garrington (\cite{aku3}), L\"udke et al. (\cite{lud}).

The basic morphology is that of a slightly bent ($\approx 20^\circ$) ``triple''
source (Fig.~\ref{evnM:6:298_a}) with a compact component dominating at 6 cm
which contains the
source core, and two extended lobes on either side of it. These are
asymmetric in ``hot--spot'' luminosity
(1.9:1 at 1.7 GHz), arm ratio (2.7:1 as measured from the ``hot--spots") and
polarization,  the Western lobe having the brighter ``hot--spot'',
being closer to the core and  unpolarized. The Eastern lobe is 
connected to the ``Central'' component by a narrow wiggling jet.

The polarization images at 5 GHz (Akujor et al. \cite{aku1}), 8.4 GHz
(Akujor \&  Garrington, \cite{aku3}) and 15 GHz (van Breugel et al.
\cite{vanb})
indicate that there is little Faraday rotation and depolarization on
the East side. The magnetic field in the jet runs parallel to its axis and is 
circumferential in the lobe. On the western side of the source no 
significant  polarization is present at any frequency, 

VLBI images of the whole structure have been produced by Graham \& Matveyenko
 (\cite{grah}) at 1.7 GHz, Nan  et al. (\cite{nan2}) at 610 MHz and Dallacasa et al.
(\cite{dalla})  at 327 MHz.
The core region has been observed with the VLBA by
Fey \& Charlot (\cite{fey}) at 2.3 and 8.5 GHz.

The 327 MHz VLBI image (100$\times$ 35 mas), reproduced from Dallacasa et al.
(\cite{dalla}), is shown
in Fig.~\ref{evn:90:298} and displays two wide tails or ``plumes'' emerging from
the extremities of each lobe; the source has then an ``S--shaped''
appearance. In this image  $\approx 30\%$ of the total flux density 
is missing, very likely in the extended  components.

\begin{figure}[h]
 \vskip 6truecm
\includegraphics{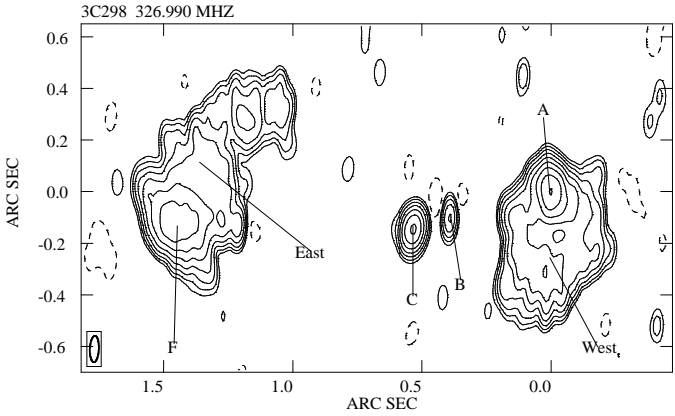}
 \caption{3C 298: EVN image at 92 cm at 100$\times$35 mas; contours: ($\pm 6.0\times 2^n, n\geq 0)$
mJy/beam; $S_{\rm peak}=$1.577 Jy/beam (from Dallacasa et al. \cite{dalla},
with
permission of the authors)}
 \label{evn:90:298}
 \end{figure}

\begin{table}[htbp]
\begin{center}
\caption{Image characteristics for 3C 298 \hfill}
\medskip
\begin{tabular}{l|ccc|ccc}
\hline
& \multicolumn {3}{c|}{1.7 GHz} &  \multicolumn {3}{c}{5 GHz} \\ \hline
resol. &  beam       & Flux$^\ddag$ & noise$^{\dag}$  &  beam       & Flux$^\ddag$ & 
noise$^{\dag}$  \\
       &  (mas)      &  (Jy)   &(mJy/b)& (mas)       &  (Jy)   &(mJy/b)\\
\hline
{\it  high}  &11$\times$5  &  4.90   &  0.64  &11$\times$5  &   1.37      &  0.25  \\
{\it int.}&26$\times$15 &  4.74   &  0.93  &26$\times$15 &   1.35      &  0.42  \\
{\it  low}   &88$\times$39 &  5.18   &  3.33  &88$\times$39 &   1.40      &  0.50  \\
\hline
\end{tabular}
\label{char_298}
\end{center}

$\dag$ computed far from the source

$\ddag$ flux density in the image
\end{table}
In Figs.  \ref{evnM:6:298_a} to \ref{core:6:298} we present a set of images  
at  1.7 GHz and 5 GHz made with different resolutions, in order to highlight 
the various features (Table~\ref{char_298}).  Images in each pair 
have been reconstructed with the same restoring beam in order to make 
the comparison of the morphologies and the calculation of the component spectra easier 
(Sect.~\ref{spec_298}).

\begin{figure}[h]
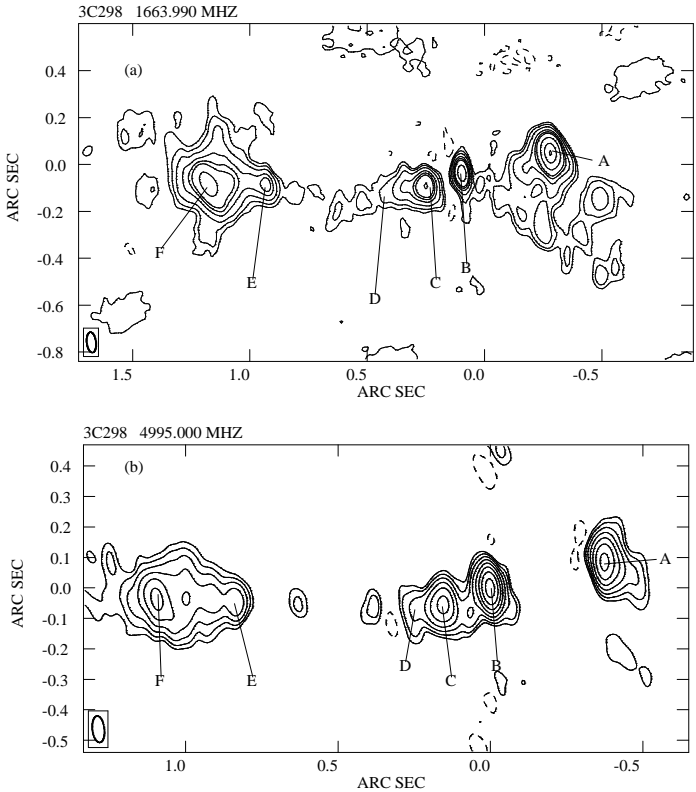

 \vskip 11truecm
\includegraphics{H3874F6A.PS}
\includegraphics{H3874F6B.PS}
 \caption{3C 298: EVN+MERLIN at 88$\times$39 mas ({\it low} resolution);  
 {\it (a)} 
 image at 18 cm; contours: (-5, $ 6.0\times 2^n, n\geq 0)$ mJy/beam;
 $S_{\rm peak}=$0.595 Jy/beam; {\it (b)} image at 6 cm;
 contours: (-3.0, $ 2.0\times 2^n, n\geq 0)$ mJy/beam; $S_{\rm peak}=$0.360 Jy/beam   }
 \label{evnM:6:298_a}
 \end{figure}

\begin{figure}[h]
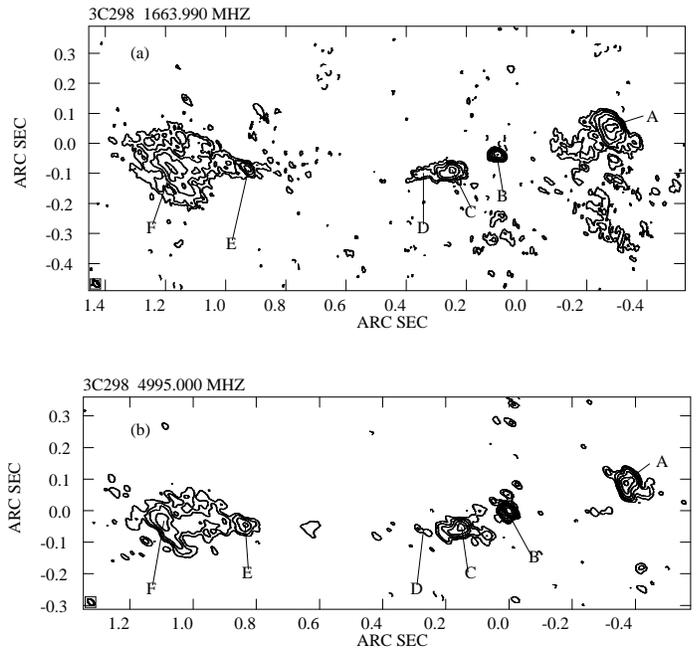

\vskip 9truecm
\includegraphics{H3874F7A.PS}
\includegraphics{H3874F7B.PS}
 \caption{3C 298: EVN+MERLIN at 26$\times$15 mas ({\it intermediate}) resolution;
{\it (a)} image at 18 cm  contours: ($\pm 3.5\times 2^n, n\geq 0)$
mJy/beam; $S_{\rm peak}=$0.229
 Jy/beam; {\it (b)}  image at 6 cm;  contours: ($\pm 1.5\times 2^n, 
 n\geq 0)$ mJy/beam; $S_{\rm peak}=$0.335  Jy/beam   }
 \label{evnM:6:298_b}
 \end{figure}

In the discussion of the source morphology we refer to the labelling of 
Fig.~\ref{evnM:6:298_a}a and Fig.~\ref{evn:90:298}, where $B$ is the ``nucleus'', $C+D$ the bright
 portion of the Eastern jet closest to the nucleus, referred to as
 {\it intermediate jet},
$E$ is the easternmost portion of the Eastern jet, $A$ and  
$F$ are the brightest regions within the two lobes,  or  ``hot--spots'', 
$EAST$ and $WEST$ refer to the extended emission {\it underlying} them. 

  At the {\it low} resolution of 88$\times$39 mas the structure seen in the
 EVN+MERLIN image at 18 cm (Fig.~\ref{evnM:6:298_a}a) closely resembles the
5 GHz  MERLIN image
by L\"udke et al. \cite{lud}. Remnants of the ``plumes'' seen in the 327 MHz image
are clearly visible especially in the Western lobe. They mostly
disappear at 6 cm (Fig.~\ref{evnM:6:298_a}b), where the two lobes are
dominated by the ``hot--spots''.
The jet to the East is clearly visible and well collimated, although the
{\it intermediate jet} ($C+D$) is shorter at 6 cm. It is initially aligned
with $B$ and $A$, then it changes direction
by $\approx 20^\circ$ to North at about 0.35 arcsec from the
nucleus, where it also becomes very faint. It brightens again at $E$, where it
meets the Eastern lobe.
On the Western side the presence of a jet is not obvious. It may be
the narrow feature on the East of $A$, which however is not visible at 6 cm.
We note that the ``jetted side'' of the source corresponds to the lobe which
is farther from the core and more polarized (L\"udke et al. \cite{lud}).

Component $B$ is brighter at 6 cm, indicating an inverted spectrum, and hence the
location of the source core (Nan et al. \cite{nan2}; van Breugel et al.
\cite{vanb}). 
In both images the measured total flux density coincides, within the errors, 
with the lower resolution measurements at these frequencies.

\begin{figure*}
 \vskip 11truecm
\includegraphics{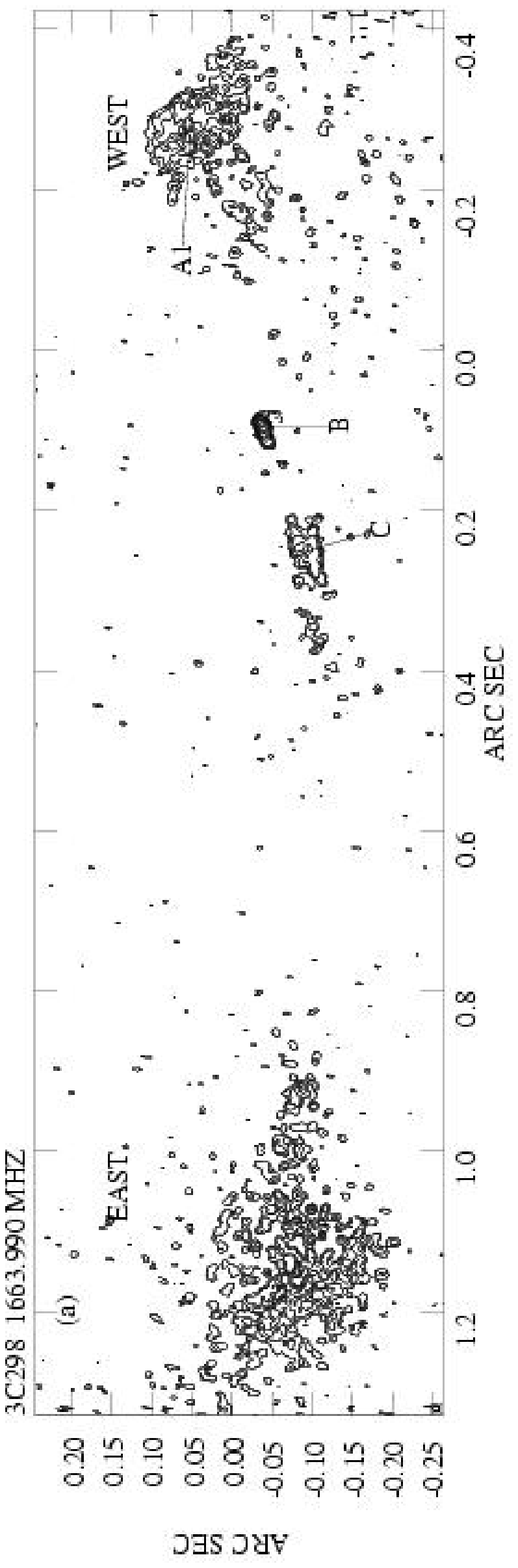}
\includegraphics{H3874F8B.PS}
 \caption{3C 298 at 8$\times$8   mas (almost full resolution): {\it (a)} 
 18 cm image;  contours: (-4.0, 
 $ 2.0\times 2^n, n\geq 0)$mJy/beam; $S_{\rm peak}=$0.156 Jy/beam;
{\it (b)} 6 cm image; contours: (-2.0, $ 1.0\times 2^n, n\geq 0)$mJy/beam; 
 $S_{\rm peak}=$0.168 Jy/beam    }
 \label{ejet:6:298}
 \end{figure*}

\begin{figure*}
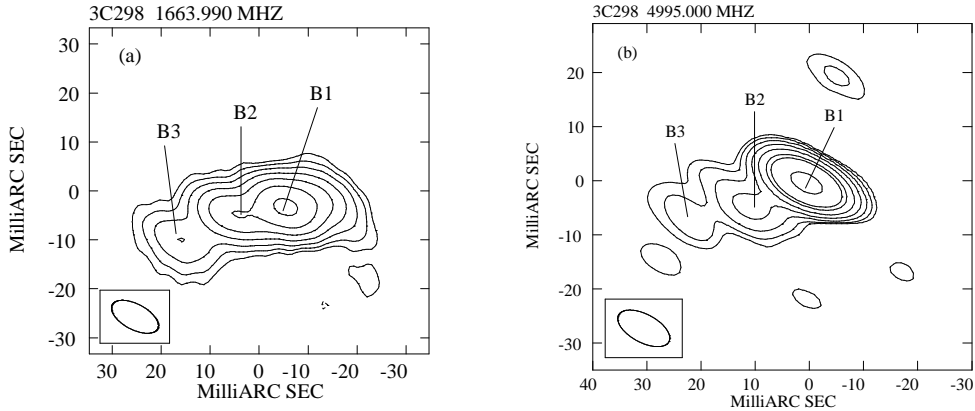

 \vskip 7truecm
\includegraphics{H3874F9A.PS}
\includegraphics{H3874F9B.PS}
 \caption{3C 298 ``core'' region at 11$\times$5  mas ({\it high}) resolution; 
 {\it (a)} 18 cm image;  contours: (-4.0,
 $ 2.0\times 2^n, n\geq 0)$mJy/beam; $S_{\rm peak}=$0.147 Jy/beam;
{\it (b)} 6 cm image;   contours: (-4.0,
 $ 2.0\times 2^n, n\geq 0)$mJy/beam; $S_{\rm peak}=$0.332 Jy/beam    }
 \label{core:6:298}
 \end{figure*}

At the {\it intermediate} resolution of 26$\times$15 mas (Fig.~\ref{evnM:6:298_b})
further extended
structure disappears into  the noise. A number of features are better delineated and
component $B$ begins to show hints of extension.
At 18 cm (Fig.~\ref{evnM:6:298_b}a) the Eastern lobe is very fragmented and
the jet at $E$, now well collimated, can be traced, on colour images (not
shown), within
the lobe. The
{\it intermediate jet} ($C+D$) is still present, quite collimated but
shorter. A ridge of emission seems to run from the ``hot--spot'' $F$ to South--West.
Of the Western lobe only the bright component $A$, now well resolved,
and hints of the plume are still
visible. The elongated feature East of $A$, that in Fig.~\ref{evnM:6:298_a}a
could have been interpreted as the Western jet end, appears now quite wide. Note that
$A$ is elongated roughly perpendicular to the jet overall direction. 
At 6 cm (Fig.~\ref{evnM:6:298_b}b) only
components $A$, $B$, $C$, $E$ and  $F$  are still clearly visible. 

Images of the entire source, restored with a circular Gaussian beam of $8\times 8$ mas,
are shown in Fig.~\ref{ejet:6:298}. Here  we see the ``nucleus'' and the start 
at $B$ of the Eastern jet ({\it near jet}), the {\it intermediate jet} at $C$ 
and, at 18 cm only, lobes $WEST$ and $EAST$, although much fragmented. 
It is interesting to note that the {\it intermediate 
 jet} has a  sharp edge on its western side at both frequencies (although
 not easy to see in Fig.~\ref{ejet:6:298}) roughly perpendicular to the jet 
 axis.  This  feature could be the result of a transverse shock. 

At full resolution ($11\times 5$ mas, images not shown) only the region $B$ to $C$ can be  well 
imaged at both frequencies. The two lobes are almost completely resolved out.
In the Western one a quite compact bright ``true'' hot--spot
($A1$ in Table~\ref{com298}), accounting for  $\approx$ 10\% of component
$A$ flux density, stands out at both frequencies clearly distinct from the 
surrounding low brightness emission. This hot--spot is barely
distinguishable from the rest of the lobe in Fig.~\ref{ejet:6:298} 
due to the compressed angular scale. The ``hot--spot'' $F$ in the Eastern lobe is 
completely resolved out at both frequencies. 

The nuclear region ($B$) is shown in Fig. \ref{core:6:298} at the highest 
available resolution. At least 3 components, labelled $B1, B2, B3$
from West to East, are visible. The VLBA image by Fey \&
Charlot (\cite{fey}) at 2.3 GHz, is in reasonable agreement with ours at 1.7 GHz,
which has a similar resolution. From the images in the 1.7--8.4 GHz range
we conclude that $B1$ is most
likely the ``true'' core in the source, since it is the most compact feature and has 
a convex spectrum peaking between 2.3 and 5 GHz (Sect. \ref{spec_298}).
Components $B2$ and $B3$ represent the {\it near jet}.

Component parameters are given in Table~\ref{com298} at both 1.7
(first line) and 5 GHz (second line). Observed parameters of all components
(but $B$)  are measured on the {\it low} resolution images 
(Fig.~\ref{evnM:6:298_a}); parameters of  $B1, B2,
B3$ and of the hot--spot $A1$ are from the {\it high} resolution data.
The flux density of the extended structure in the two lobes
has been determined by measuring the whole lobe flux density with task 
TVSTAT (Sect.~\ref{data_an}) and by then subtracting the contribution of the
bright components $A$ and $F$ respectively. Their sizes have been estimated 
from  the contour plots. For all the other components flux density and
beam--deconvolved size were obtained using IMFIT.

Derived physical parameters for all components but $B1$ (which is unresolved and
with a convex spectrum) have been computed assuming equipartition conditions
and are reported in Table~\ref{com298}. They are computed from the 1.7 GHz
data, except for $B2$ where we used the 5 GHz data.

\begin{table*}
\begin{center}
\caption{Observed and Derived Parameters for 3C 298\hfill}
\smallskip
\begin{tabular}{c|rcc|rrrrr}
\hline
comp & S & $\theta_1\times\theta_2$ &  d$_1\times$d$_2$ &
$\alpha_{\rm thin}$ & ~~H$_{\rm eq}\ \ $ & u$_{\rm min}\ \ \ $ & U$_{\rm eq}\ \ \ \ $ &
$\nu_{\rm to}$\\
       & mJy   &     mas                 &     pc $~h^{-1}$   &        &
                     ~~ mG$\ \ $          & erg cm$^{-3}~h\ \ $  & 10$^{54}$erg $h^{-2}$   &
   MHz \\
\hline
$A$      & 1024 &  57$\times$45 &  245$\times$193&1.28 &4.3~~~&1.8$\times
10^{-6}$ & 1400~~~~~&234 \\
         &  193 &39$\times$24  &                &     &    & & & \\

$\dag A1$ & 63 &$\ltsim 2$&$\ltsim 9$ &0.49 &$\gtsim$15~~
&$\gtsim$2$\times10^{-5}$  &$\ltsim$1~~~~~ &$\gtsim$713 \\
(h.sp)&   37 &$\ltsim 2$  & & & & & & \\

$\dag B1$    &  254 &9$\times$2 &26$\times$9 &0.82 & & & & $\sim$3000\\
 $    $      &   330 &unres  & & & & & & \\

$\dag B2$      &   51 &unres &  &0.80  &  &   & & \\
       &   29 & 5$\times$4   &21$\times$17 & &1.0~~~ &1$\times10^{-5}$ &5.6~~~~~ &477 \\

$\dag B3$      &   33 &9$\times$6 &39$\times$26&0.87 &5.5~~~
&2.8$\times10^{-6}$ &6.4~~~~~ &238 \\
       &    9 &    & & & & & & \\

$C$    &    245 &  43$\times$14 & 185$\times$60&0.74 &3.4~~~ &1.1$\times
10^{-6}$ &  62~~~~~&187 \\
       &    102 &  41$\times$16                &     &    & & & &\\

$D$    &    292 & 157$\times$58 & 674$\times$249&1.20 &1.8~~~~&
3.0$\times 10^{-7}$ & 1100~~~~~& 103 \\
       &     77 & 162$\times$28 &                 &     &    & & & \\

$E$    &    183 &  72$\times$26 &  309$\times$112&1.50&4.3~~~ &
1.7$\times 10^{-6}$ & 590~~~~~& 173 \\
       &     36 &  $<85\times35$ &                &     &    & & & \\

$F$    &  416 & 108$\times$32 & 463$\times$137&0.78 &1.9~~~ &
3.6$\times 10^{-7}$ & 275~~~~~&118\\
   &  107 & 54$\times$30 &                 &     &    & & & \\
\hline
$WEST$     & 973 & $\sim$700$\times$400 & (3.0$\times1.7)\times 10^3$&1.50
&0.6~~~~& 0.3$\times 10^{-7}$ &3900~~~~~& 75\\
           &  87 & $\sim$470$\times$320 &                &     &    & & & \\

$EAST$     &  1810 &$\sim$800$\times$400 & (3.4$\times1.7)\times 10^3$&1.03
&0.6~~~~& 0.6$\times 10^{-7}$ &4400~~~~~& 70\\
           &  346 & $\sim$830$\times$470 &                &     &    & & & \\
\hline
\end{tabular}
\label{com298}
\end{center}
1st line: 18 cm data; 2nd line: 6 cm data

$\dag$ data from the {\it high} resolution image; flux density of $A1$ included
in  $A$

$\alpha_{\rm thin}$ spectral index in the frequency range 0.3--5 GHz ;
H$_{\rm eq}$ equipartition magnetic field; u$_{\rm min}$ minimum energy
density; U$_{\rm eq}$ minimum energy; $\nu_{\rm to}$ computed turnover
frequency (except for $B1$ which is observed; Sect.~\ref{spec_298})

\end{table*}

\subsubsection {Jet brightness distribution}
\label{j-bri}

In the Eastern jet ({\it near} and {\it intermediate} portions) we have 
analyzed the surface brightness $B(\phi)$ at 1.7  GHz as a function of the
(beam--deconvolved) jet transverse
size ($\phi$) measuring the HPW of the jet perpendicular to its direction
at several positions spaced by about one beam, in such a way as to have
independent measurements.

The average jet opening angle is  $\approx 17^\circ$.
The deconvolved brightness decreases as
$B(\phi)\propto\phi^{-m}$ with $m$ in the range $0.5-2$. In spite of the
uncertainties, we conclude that the jet appears highly sub--adiabatic.

\subsubsection{ Spectral Analysis}
\label{spec_298}

The overall spectrum of 3C 298, derived from low resolution measurements (Kuhr et al.
\cite{kuh}; Kameno et al. \cite{kame}; Steppe et al. \cite{step}, Murgia et al. 
\cite{mur})
 is straight and steep ($\alpha=$1.15) from $\approx$90 GHz (or possibly
 230 GHz) down to $\approx$80 MHz, where it turns over.

An attempt to analyze the multi--frequency spectra (from 0.3 to 22 GHz) of the
individual components has been carried out adopting a ``low--resolution approach''
(except for $A1$ and
subcomponents of $B$ for which we used the data at the maximum available
resolution). A complication is that some components are blended at
some frequencies and well separated at others, so that the spectral indices,
$\alpha_{\rm thin}$ in Table~\ref{com298}, do not always make use of all the
six available frequencies (0.3, 0.6, 1.7, 5, 15, 22 GHz, 
Sect.~\ref{mor_298}).
In addition, in order to fully exploit the whole frequency range,
we have also fitted the spectra of the blends $E+F$ and $B+C+D$.
For components $A1, B1, B2, B3$ two to four frequencies, from this 
paper and from Fey \& Charlot (\cite{fey}), are available. 
All components (and blends), except  $B1$, have straight spectra in
the available frequency range (often down to 0.3 GHz). Component $B1$,
instead, has an inverted spectrum with a maximum around 3 GHz and 
$\alpha_{\rm thick}$ close to 2.5 between 1.7 and 2.3 GHz.

The two lobes $EAST$ and $WEST$ are heavily resolved in the 15 and 22 GHz 
images of van Breugel et al. (\cite{vanb}) as well as at the low frequencies of 0.6 
and 0.3 GHz. Therefore the spectral indices in Table~\ref{com298} 
are computed using only the present flux densities at 1.7 and 5 GHz.

As done in Sect.~\ref{spec_43} we computed the source {\it subtracted spectrum}
by subtraction of the (core $+$ jet) flux densities
(components $B + C\&D$) from the overall spectrum. This is a good
estimate of the global spectrum of the two lobes (``hot--spots'' 
included) and of the eventually missed low surface brightness features.
This spectrum shows a break at $\approx$ 1 GHz, with spectral indices
$\alpha_{\rm low} \approx 1$ and $\alpha_{\rm high} \approx 1.6$ 
respectively below and above it.

\section{Discussion}
\label{disc}

\subsection{Sources'  Morphology}
\label{morph}

In spite of the similarity in radio power the two sources are very dissimilar 
in radio morphology. 
3C~298 has the typical  FRII characteristics of a quasar: well separated lobes
with ``hot--spots'', a one sided jet and a bright core. Most of the radio
emission at $\nu \leq 2$ GHz is from the lobes and the ``hot--spots''.

On the other hand, 3C 43 has its radio emission dominated by a {\it one-sided}
jet with sharp bends. Morphologies like
these are seen among CSSs (see Mantovani et al. \cite{manto} for a 
collection of similar objects), although they are not the majority.
{\it No} bright features such as {\it hot--spots} are
seen in the outer broad components. The overall structure is then far from an 
FRII type. 

 \subsection{Relativistic Effects and Source Orientation}
 \label{doppler}

The presence of relativistic effects in the core and jets of the sources 
are evaluated by analyzing the {\it core dominance} and the { \it jet 
asymmetry}.

\smallskip
The {\it core dominance} at 5 GHz is the ratio $R_{\rm c}=S_{\rm c}/S_{\rm ext}$ of the $k-$corrected
flux density in the core to that in the extended features.
We assume that what we call  ``core'' is actually the sum of the advancing and
of the receding bases of the jet, moving on both sides of the
true core  at the same speed ($\pm\beta_{\rm c}$) and at the same angle
($\theta_{\rm c}$) to the line of sight. 
The value found for $R_{\rm c}$ is then compared with the median value
$\left<R_{\rm c}\right>=$0.05$\pm$0.03 (error is 2$\sigma$) found at 5 GHz by Fanti et al.
(\cite{fan3}) for
CSS quasars of similar radio luminosity. $\left<R_{\rm c}\right>$ is related to the
median  angle that quasars make to the  line of sight
($\left<\theta\right>\approx 30 ^\circ$ in the {\it unified  scheme}, Urry \& Padovani,
\cite{urry}).
From the ``normalized core power''  defined as $P_{\rm c,n} =
R_{\rm c}/\left<R_{\rm c}\right> \approx \left[(1 - \beta{\rm _c}~
\cos\left<\theta\right>)/(1-\beta_{\rm c}~
\cos\theta_{\rm c})\right]^{(2+\alpha_{\rm c})}$ (Giovannini et al. \cite{giov}) one can 
estimate\footnote{To
be more precise, the adopted value of $\left<R_{\rm c}\right>$ refers to {\it steep spectrum} quasars,
which are likely oriented at angles slightly larger than average.
Since however  $P_{\rm c,n}$ depends on $\left<\theta\right>$ via a cosine we did not 
consider this a serious bias.} ($\beta_{\rm c} ~\cos \theta_{\rm c}$). 
Note that in the calculations we have ignored the contribution of the 
receding jet, since it is negligible, and have used
$\alpha_{\rm c}=0$ to be consistent with Fanti et al. (\cite{fan3}).

\smallskip
The {\it jet/counter--jet} brightness ratio is given by:

$R_{\rm j}=\left(\frac{\displaystyle{{1+\beta_{\rm j}\cos\theta_{\rm j}}}}
                      {\displaystyle{{1-\beta_{\rm j}\cos\theta_{\rm j}}}}
		      \right)^{2+\alpha_{\rm j}}$

\noindent
where $\beta_{\rm j}$ and $\theta_{\rm j}$ are the jet and counter--jet speed and the
angle
to the line of sight (both assumed to be identical for the two sides of the
source) and $\alpha_{\rm j}$ the jet average spectral index.
The most appropriate images for this purpose would be those at 1.7 GHz, which 
however do not show any obvious counter--jet.

\smallskip\noindent
\underline{\it Core Dominance}

The two sources are quite different in {\it core dominance}, since we find:

$~~~~~~~R_{\rm c}=0.012~~~~~  P_{\rm c,n} = 0.24^{+0.35}_{-0.9} ~~{\rm for}~~~
3{\rm C}~ 43$

$~~~~~~~R_{\rm c}=0.103~~~~~  P_{\rm c,n} = 2.1^{+3.1}_{-0.8} ~~~~~{\rm for}~~~
3{\rm C}~ 298$

\noindent
(errors are 2 $\sigma$ and are mainly due to the uncertainty in
$\left<R_{\rm c}\right>$).

If we assume, according to the {\it unified scheme}, that quasar jets are
oriented closer than $\approx 45^\circ$ to the line of sight, 
the ranges of  the possible angles and $\beta_{\rm c}$ become:

$35^\circ\ltsim\theta_{\rm c}\ltsim45^\circ ~~~~\beta_{\rm c}\gtsim 0.75~~~~~$ for 3C 43 

$~~~~~~~~~\theta_{\rm c}\ltsim30^\circ ~~~\beta_{\rm c}\gtsim 0.7~~~~~~~$ for 3C 298

\medskip\noindent
\underline {\it Jet/counter--jet ratio}

In neither source do we detect the counter-jet, so that we can only set
upper limits.
To be conservative and to minimize the effects of individual bright knots,
we used the average surface brightness on the jet side and twice the
local average r.m.s. noise, as upper limit, for the undetected counter--jet.
We find:

$~~~~~~R_{\rm j} \gtsim 15~~~~~~~~~$ for 3C 43  \ \ -----$> \beta_{\rm j}
~\cos\theta_ {\rm j}\gtsim 0.46$

$~~~~~~~~~~\gtsim 5~~~~~~~~~~$ for 3C 298 -----$> \beta_{\rm j} ~\cos\theta_{\rm j} \gtsim 0.3$

The above limits do not constrain the jet orientation and speed any better
than the {\it core dominance}. We note however that for the jet of 3C~43
lower speeds and smaller angles to the line of sight than derived from the
{\it core dominance} would be permitted, but this would imply a change in the
jet direction out of the core $A$ (see also Sect.~\ref{dist3c43}).

\subsection{Distortions}
\label{dist3c43}

3C 43 appears as a very distorted radio source, with sharp bends:
$\approx 40^\circ$ at $C2,~ \approx 60^\circ$ at $D$ (\app 200 mas, i.e. \app
850 pc$~h^{-1}$), $\approx 30^\circ$ at
$F$ (see Figs. \ref{new_M}b and \ref{evnM:18:43}). 
Note that all these sharp bends appear to occur where a bright knot of 
emission is seen, suggestive of jet--ISM interactions.

On the basis of the present
 knowledge of NLR properties, Mantovani et al. (\cite{manto}) estimated that 
 in up  to 10\% of CSSs the jets are       
 likely to hit a dense cloud  which  deflects them  without  disruption.
The physics of jet-cloud interaction has been investigated by de Young
(\cite{dey}) and Norman \& Balsara (\cite{norm}) in 3-D hydrodynamical simulations.
They show that a jet may maintain its collimation for deflections up to
90$^\circ$. So the distortions we see may well be due to jet interactions with
inhomogeneities of the ambient medium. 
 
Sharp deflections might also be due to projection effects on a 
moderately distorted jet seen close to the line of sight. But this does not 
seem to be the case for 3C~43. According to Eq. (\ref{proj}) in the Appendix,
the large bends
we see could only be produced by projection effects if the  {\it
bright jet} is oriented at a very small angle to the line of sight
(see also the extended discussion in Conway \& Murphy \cite{con}).
This however is somewhat in contradiction with the discussion  of 
Sect.~\ref{doppler},
where, from the relative weakness of the core, angles larger than $\sim
30^\circ$ were suggested which  are too large to produce large apparent
deflections. For instance, for $\theta\approx 30^\circ$, an observed bend
$\zeta~'\sim 95^\circ$ is obtained only with an intrinsic bend
$\zeta\gtsim 40^\circ$.
Of course we cannot exclude that the mini-jet within $A$ (Sect. 
\ref{mor_43}) be at large angles to the line of
sight  (as deduced from the {\it core dominance}), and that it changes
its orientation at component $B$. The {\it bright} and {\it faint jet} 
(Fig.~\ref{evnM:18:43}) would then be an intrinsically almost straight jet, oriented 
close to the line of sight (in agreement with the lack of a counter--jet,
Sect.~\ref{doppler}), whose visible bend is due solely to projection
effects. 
But this would represent again a large intrinsic distortion occurring
between $A$ and~$B$. So it seems to us not very
plausible that all the large bends we see are amplifications of small ones.

We note, finally,  that all the bends are always in the same sense, as, e.g.,
in 3C~119 (Nan et al.  \cite{nan1}) and in 3C~287 (Fanti et al. \cite{fan2}). 
Therefore it appears
 unlikely that  they are just due to {\it random strikes} of the jet 
 against several dense NL clouds. A mechanism which governs on which side the jet
has to turn around seems to be required (see discussion in Nan et al.
\cite{nan1}). An  alternative possibility is that we
are seeing a helical jet in  projection (Conway \& Murphy \cite{con}), 
but this seems implausible since this model applies to high $\gamma$
core dominated objects, while 3C~43 it is not.

\smallskip
3C 298 has a much more linear structure, compared to 3C 43.
Figs. \ref{evnM:6:298_a} to \ref{ejet:6:298} show a
gentle regular bending. ``Hot--spot'' $A$ is in p.a.\app$-73^\circ$ with respect
to $B$. The Eastern jet starts with p.a.\app $120^\circ$ and is roughly
aligned with the {\it intermediate  jet} (Fig.~\ref{ejet:6:298}) then it
deviates northward by \app $20^\circ$ at about 350 mas (or 1.5 kpc $~h^{-1}$). Since in
this case the jet is plausibly oriented at small angles to the line of sight,
it is likely that a small intrinsic bending is amplified to the observed value
by projection effects.

\subsection{Source Ages}
\label {s-age}
A conventional way to estimate a source age, with all the necessary caveats, is
via its radiative age. To do so we estimated for both sources the radiation
loss frequency break, $\nu_{\rm break}$, in the {\it subtracted spectrum} 
(Sect.~\ref{spec_43} and \ref{spec_298}), which we consider a good 
approximation of the overall spectrum of the more extended, 
and hence plausibly older, components, visible or not in the present 
observations. As pointed out by Murgia et
al. (\cite{mur}), radiative ages are likely to represent the source age only
when the lobes, which have accumulated the electrons produced over the source
lifetime, dominate the source spectrum, as it is the case for 3C~43 and 3C~298.

Both {\it subtracted spectra} may be fitted by a Continuum Injection model.
From $\nu_{\rm break}$ the spectral age has been estimated adopting the
equipartition magnetic field (H$_{\rm eq }$). Such ages have been compared, 
whenever possible, with estimates obtained with different methods.

In 3C 43 the $\nu_{\rm break}$ in the {\it subtracted spectrum} falls in the range
$\approx 0.3$ GHz to $\ltsim
0.1$ GHz, depending on the low frequency behaviour of spectrum of the
``Central'' component. For an estimated equipartition magnetic field
H$_{\rm eq }\approx 
0.5 $ mG, the radiative {\it  age} of the extended components
is in the range 2 to 3 $\times 10^5$ years.

In the case of 3C~298 the {\it subtracted spectrum} has
$\nu_{\rm break}\approx  1$  GHz. 
This implies, for the estimated equipartition magnetic field H$_{\rm
eq}\approx0.6$ 
mG, a {\it radiative age} $\tau_{\rm r}\approx 7~10^4$y for the
extended components.

The age estimates for both sources disagree somewhat with those 
derived by
Murgia et al. (\cite{mur}). This is just due to the different frequency breaks adopted 
by those authors for the total spectrum, which is affected by the presence of 
the compact structures, and to the equipartition magnetic fields, poorly
estimated due to the lack, at that time, of high resolution images. 

\smallskip
For 3C 43, which is very twisted and with no visible
``hot--spots'' (Sect.~\ref{morph}), we have no other way to estimate its age.
For 3C 298 instead, we have two alternative approaches.

\noindent
{\it (a)} -- We noticed earlier that the ``jetted lobe''
is significantly farther from the core than the
``un-jetted'' one. This is expected if the heads of the lobes advance
at a velocity ($\beta_{\rm h}$) high enough that there are different travel time 
delays for the radiation from them.
The more distant lobe would be the one advancing towards the observer. This is
confirmed by the asymmetry in the polarization of the two lobes (Akujor
\& Garrington, \cite{aku3}), if the latter is interpreted in terms of the
``Laing--Garrington'' effect (Laing, \cite{lai}; Garrington et al. \cite{garr}).
The observed {\it arm ratio} is $R_{\rm arm} \approx 2.7:1$.
On the assumption that the arm asymmetry is solely due to travel time delays,
one obtains a $\beta_{\rm h}\cos \theta \approx 0.45$. 
For any $ \theta \ltsim 30^\circ$ (as in the core, Sect.~\ref{doppler}) we have 
$\beta_{\rm h} \approx 0.45-0.5$.
The above value of $\beta_{\rm h} \cos\theta$ would imply
a luminosity ratio of the two ``hot--spot'' $R_{\rm h.sp.} = R_{\rm arm}^{3+\alpha}
\approx 40$, while the observed $R_{\rm h.sp}.$ is $\approx 2.4$ only, and 
reversed, the supposedly receding ``hot--spot'' being more luminous than the 
approaching one. Part of the discrepancy may be attributed to the fact that the receding
``hot--spot'' $A$ is seen at an earlier stage of evolution compared to the
advancing one $F$, which may have suffered radiative/adiabatic losses,
but this is not sufficient.
We could also speculate that the western arm is shortened by projection
if the source structure is not linear but more bent toward the line of sight
on the Western side, or that the ``hot--spot'' luminosities are dominated by
relativistic back--flows.  However, if we ignore these contradictions and 
assume that the lobe arm asymmetry is due to travel time delay,
we deduce a lower limit for the ``kinematic age'' of the advancing and of the
receding lobes of $\approx 3.2 \times 10^4 \times (\tan \theta)^{-1}$ years
and $\approx 1.1 \times 10^4 \times (\tan \theta)^{-1}$ years respectively, which, 
for $\theta \leq 30 ^\circ$, are very close to the radiative age estimate.

\noindent
{\it (b)} --  We can estimate of the source age from {\it energy budget}
arguments. The {\it equipartition pressure} in the ``hot--spots''
($p_{\rm h.sp.}$) allows us to compute the {\it jet thrust} defined as
$\Pi \approx p_{\rm h.sp.} \times {\cal A}$, where ${\cal A}$ is the jet impact
area (estimated from the ``hot--spot'' diameters), and the {\it jet energy flux}
is defined as
$F_{\rm e,j} = c ~\Pi $. The time required to feed the lobes ({\it feeding age})
is then derived from the ratio $2 \times U_{\rm lobe}/F_{\rm e,j}$, where
$U_{\rm lobe}$ is the lobe ($EAST$ or $WEST$) minimum total energy and
the factor 2 roughly accounts for the work spent to expand the lobe. We
obtain 
{\it feeding ages} $\tau_{\rm f} \approx 2-5 \times 10^4$ years for $WEST$
and $EAST$ respectively, in fair agreement
with the previous estimates.

\subsection{The external medium}
\label{extmed}

Some inferences on the properties of  the medium surrounding the two sources
can be obtained from their physical parameters.
In the case of 3C~298, taking the source growth velocity derived from the
arm ratio in Sect.~\ref{s-age}, the balance between the  {\it ram pressure}
and the
``hot--spot'' pressure allows us to estimate the external medium density. We obtain
$n_{\rm e} \approx 1 \times 10^{-3}$~cm$^{-3}$ for ``hot--spot'' $A$ and a value two
times lower for ``hot--spot'' $F$. Such a density  estimate
is quite low with respect to others found in the literature (see 
e.g. O'Dea, \cite{ode} and references therein).

Taken at face value, these figures would
indicate a decrease of the density with the distance $r$ from the core as
$n_{\rm e}
\propto r^{-0.7}$, suggesting, from the value of the exponent of $r$,
that the lobes are just crossing the gas core radius. 
The internal pressure of the lobes, coupled to the above density 
estimate, is incompatible with a static confinement even for very high 
gas temperatures. It is therefore very likely that
the lobes are {\it over-pressured} and therefore that they are expanding
supersonically.

As pointed out in Sect.~\ref{morph}, 3C 43 lacks  ``hot--spots''. We 
remark, however,
that its broad components have internal pressures not too far from 
those of the lobes of
3C~298. We are tempted to assume that the external medium properties 
and dynamics of expansion of the broad components are similar in the two
sources. However the medium has to be very clumpy in this source, 
if the bends are caused by
jet-cloud interaction. Furthermore a special space distribution of the
clouds is required, in order to act as the {\it wall of a cavity} (see 
Nan et al. \cite{nan1}) and to cause the jet to bend always in the same sense.

\section{Summary and conclusions}
\label{summ}

In this paper we have presented multi--frequency and multi--resolution
observations of two CSS quasars from the 3CR catalogue: 3C~43 and 3C~298.

The two sources have similar redshift and radio power, but appear very
different in their properties.
3C~298 has all the characteristics of a small FRII radio source.
It shows a moderately bright radio core, a one-sided jet and lobes with
``hot--spots''. It is likely oriented at $20^\circ - 30^\circ$ to the line of
sight but its deprojected linear size is still  $\leq 20$~kpc$~h^{-1}$. Its
radio luminosity at $\nu \leq 2$ GHz is largely provided by lobes and
``hot--spots''. 

\smallskip
3C 43 is an object hard to interpret. It is dominated by a knotty jet showing
several large bends which cannot be explained by projection effects
if the overall source orientation is $\gtsim 30^\circ$, as deduced from
the weakness of its core. In order to bring the core weakness in agreement
with the large distortions being due to projection, one should assume that 
the jet changes its orientation just out of the core. This would also explain, 
via Doppler de--boosting, the fact that no counter-jet is seen, in spite of 
the core weakness. But again, if the jet turns toward the observer 
just outside the core, this may represent a large intrinsic deflection.
We are led to interpret 3C~43 as a radio source intrinsically distorted by  jet--cloud
interactions. The external medium has to be clumpier than average, in order 
to cause several large bends, and has to have a very special space
distribution to produce the overall clock-wise distortion of the source.

\smallskip
The equipartition magnetic fields are in the range 2--5 mG in most components,
reaching several tens of mG in the cores and sub--mG values in the lobes, in
agreement with Fanti et al. (\cite{fan5}).

\smallskip
The estimated radiative ages of the extended components of \app 2~10$^5$ years for 3C~43 and  of $\approx 7 \times 10^4$ 
years for 3C~298 
suggest that these two quasars are moderately young. The
radiative age of 3C298 is supported by other arguments based on
energy budget considerations and on the arm ratio of the two lobes.
The growth velocity of 3C 298 is probably $\approx 0.5 c$. 
The density of the external medium is estimated to be 
$ \approx 10^{-3}$ cm$^{-3}$.

We have no such additional arguments for 3C 43.

 \begin{acknowledgements}
We thank the referee G. Taylor for the many useful comments.
The European VLBI Network is a joint facility of European and Chinese
Radio Astronomy Institutes funded by their National Research Councils.
MERLIN  is the Multi--Element Radio Linked Interferometer Network and is a
national facility operated by the University of Manchester on behalf of
PPARC. 
The VLBA and (US network) is (was) operated by the U.S. National Radio Astronomy Observatory
which is a facility of the National Science Foundation operated under
a cooperative agreement by Associated Universities, Inc.
This research has made use of the NASA/IPAC Extragalactic Database (NED) 
which is operated by the Jet Propulsion Laboratory, California Institute of 
Technology, under contract with the National Aeronautics and Space
Administration. 
This work has been partially supported by the Italian MURST under
grant COFIN-2001-02-8773.

\end{acknowledgements}

\appendix
\section {Effect of projection on jet bends}
For convenience of the reader we summarize here the geometry of jet
projection (see also Moore et al. \cite{moor}, Conway \& Murphy \cite{con}).
Consider the right--handed coordinate system of Fig.~\ref{schizzo} with $Z$
along the line of sight and $XY$ in the plane of the sky. Suppose for
simplicity that the jet is made of just two straight segments: $Jet_1$ in the
plane $ZX$ and at an angle $\theta$ to the line of sight ($Z$), $Jet_2$
at an angle $\zeta$  with respect to $Jet_1$
($\zeta=0^\circ$ no bending,  $\zeta=180^\circ$  the jet turns
completely backward). The locus of the possible positions in space of
$Jet_2$ is the surface of a cone with axis $Jet_1$ and half opening angle
$\zeta$. The position of $Jet_2$ on the conical surface is then
identified by the azimuthal
angle $\phi$ such that for  $\phi=0^\circ$ or $\phi=180^\circ$
$Jet_2$  lies in the plane $ZX$  and for  $\phi=\pm 90^\circ$
$Jet_2$ is parallel to $ Y$.
\begin{figure}[h]
\vskip 7cm
\includegraphics{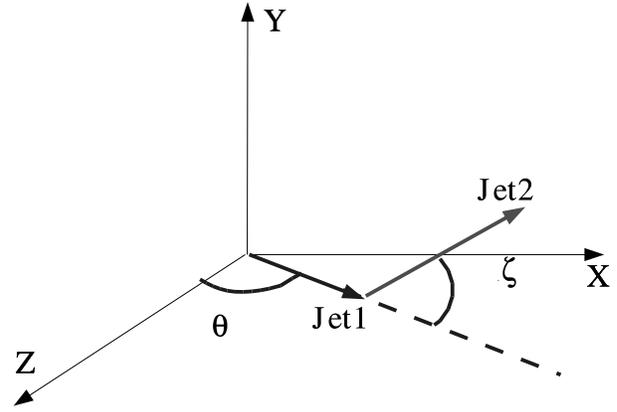}
\caption{Scheme for jet projection. $Z$ is along the line of sight; $XY$ is
the plane of the sky}
\label{schizzo}
\end{figure}
The relation between the apparent ($\zeta~'$) and the intrinsic\ ($\zeta$) 
bending angle is:

\begin{equation}
\tan\zeta~'=\frac{\sin\zeta\sin\phi}{\cos\zeta\sin\theta+\sin\zeta\cos
\theta\cos\phi}
\label{proj}
\end{equation}

It is clear from this equation that certain combinations of $\theta$ and
 $\phi$ could produce $\zeta~'\approx 90^\circ$, even for small $\zeta$.
 For instance for  $\phi=90^\circ$ $\zeta~'\tende 90^\circ$ when
 $\theta\tende 0^\circ$,  independent of $\zeta\not= 0$.

\end{document}